\documentclass[aps,prc,notitlepage,floatfix,nofootinbib,superscriptaddress,amsmath,amssymb,longbibliography]{revtex4-1}

\usepackage[utf8]{inputenc}
\usepackage[T1]{fontenc}
\usepackage{lmodern}

\usepackage{mathrsfs}
\usepackage{ulem}
\usepackage{graphicx}
\usepackage{epsfig}
\usepackage{bm}
\usepackage{color}
\usepackage{float}
\usepackage{dcolumn}
\usepackage{multirow} 
\usepackage{xcolor}
\usepackage{hyperref}
\hypersetup{
    colorlinks,
    linkcolor={red!50!black},
    citecolor={blue!50!black},
    urlcolor={blue!80!black}
}
\usepackage{color}
\usepackage{braket}
\usepackage{amsmath}

\newcommand{\be}{\begin{equation}}
\newcommand{\ee}{\end{equation}}
\newcommand{\bea}{\begin{eqnarray}}
\newcommand{\eea}{\end{eqnarray}}

\begin{document}

\title{Infrared extrapolations of quadrupole moments and transitions}

\author{D. Odell} 

\affiliation{Department of Physics and Astronomy, University of
  Tennessee, Knoxville, Tennessee 37996, USA}

\author{T. Papenbrock} 

\affiliation{Department of Physics and Astronomy, University of
  Tennessee, Knoxville, Tennessee 37996, USA}

\affiliation{Physics Division, Oak Ridge National Laboratory, Oak
  Ridge, Tennessee 37831, USA}

\author{L. Platter} 

\affiliation{Department of Physics and Astronomy, University of
  Tennessee, Knoxville, Tennessee 37996, USA}

\affiliation{Physics Division, Oak Ridge National Laboratory, Oak
  Ridge, Tennessee 37831, USA}


\begin{abstract}
  We study the convergence of bound-state quadrupole moments in finite
  harmonic oscillator spaces.  We derive an expression for the
  infrared extrapolation for the quadrupole moment of a nucleus and
  benchmark our results using different model interactions for the
  deuteron. We find good agreement between the analytically derived
  and numerically obtained convergence behavior. We also derive an
  extrapolation formula for electric quadrupole transitions and find
  good agreement with the numerical calculation of a simple system.
\end{abstract}


\maketitle

\section{Introduction}
\label{sec:introduction}
The numerical calculation of observables of strongly interacting
systems requires frequently the use of truncated Hilbert spaces. For
example, lattice quantum chromodynamics (QCD) simulations are usually
carried out in a finite volume with periodic boundary
conditions. Nuclear structure calculations employ frequently the
harmonic oscillator (HO) basis as it preserves rotational symmetry and
facilitates a straightforward way of separating out the center-of-mass
motion, see, e.g., Refs.~\cite{navratil2009,barrett2013}. Such
calculations require clearly a quantitative and qualitative
understanding of the corrections due to the involved Hilbert space
truncation, and in lattice QCD the general form of these corrections
were derived for a number of observables by L\"uscher approximately 30
years ago
\cite{luscher1985}. For nuclear structure calculations in the HO basis
it was only recently understood that the truncated HO can be thought
of as imposing long-range, hard-wall boundary conditions with an
additional short distance regulator~\cite{furnstahl2012}.
Specifically, it was found that a HO basis consisting of $N$
oscillator shells with oscillator length $b$ has an ultraviolett (UV)
cutoff~\cite{stetcu2007}
\begin{equation}
\label{Lambda_approx}
\Lambda\approx\sqrt{2N}/b~,
\end{equation}
while the infrared (IR) cutoff and therefore the spatial extent of the
basis is approximately~\cite{hagen2010b,jurgenson2011}
\begin{equation}
\label{L_approx}
L\approx \sqrt{2N}b~.
\end{equation}
Relations (\ref{Lambda_approx}) and (\ref{L_approx}) are
leading-order approximations and valid for $N\gg 1$. A more precise
expression for an HO in three dimensions was derived in Ref.~\cite{more2013}
\begin{equation}
L = \sqrt{2(N_{\rm{max}} + 3/2 + 2)}b~.
\label{L}
\end{equation}
Here, $b = \sqrt{\hbar/(\mu\Omega)}$, $\mu$, and $\Omega$ denote the
oscillator length, the reduced mass and the oscillator frequency, respectively.  We
note that Eq.~\eqref{L} is specific to a two-body system in relative
coordinates (or a single particle in three dimensions). Precise values
for the IR length scale $L$ were also derived for many-body product
spaces~\cite{furnstahl2014b}, and no-core shell model
spaces~\cite{wendt2015}.

Coon {\it et al.} \cite{coon2012} found that ground-state energies converge
exponentially with the IR length $L$. This convergence can be
understood as follows~\cite{furnstahl2012}. The finite extent $L$ of
the oscillator basis in position space imposes a Dirichlet boundary
condition of the bound-state wave function at $r=L$. The exponential
convergence in $L$ is thus directly related to the exponential fall-off
of bound-state wave functions in position space. These insights led to
theoretically founded IR extrapolation
formulas~\cite{furnstahl2012,furnstahl2014} for bound-state energies
\begin{equation}
\label{extraE}
E_L = E_\infty + a_0 e^{-2k_\infty L}~, 
\end{equation}
and radii
\begin{equation}
\label{extraR}
\braket{\hat{r}^2}_L \approx \braket{\hat{r}^2}_\infty
- \left[c_0 (k_\infty L)^3 + c_1 k_\infty L\right] e^{-2k_\infty L}~.
\end{equation}
Here $a_0$, $k_\infty$, $E_\infty$, and $c_0$, $c_1$, and $\braket{\hat{r}^2}_\infty$
are determined by fitting to numerical data in many-body systems. 

In both cases (as well as for the quadrupole moment extrapolation
derived below) the $e^{-2 k_\infty L}$ term comes from the universal
long-range behavior of the radial wave function, $\mathcal{R}_l(r)$.
The spherical Hankel functions, $h_l(\pm ik r)$, are the
negative-energy solutions in the free region, and imposing a Dirichlet
boundary condition at $r = L$ gives a solution of the form 
\begin{equation}
\mathcal{R}_l(r) = h_l(i k r) + C h_l(-i k r)~,
\label{radial_general}
\end{equation}
where $C = e^{-2 k L}$ in leading order for $k L\gg 1$.

In this work, we derive an IR extrapolation formula for the quadrupole
moment
\begin{equation}
\label{quad}
\bra{{\bf r}'}\hat{Q}\ket{\bf{r}} = 
e \sqrt{\frac{\pi}{5}} r^2 Y_{20}(\theta, \phi) \delta^{(3)}({\bf r} - {\bf r}')~,
\end{equation}
and take the deuteron as an example. For the deuteron, $r$ is the
relative coordinate. While computing the deuteron's quadrupole moment
poses no challenge in HO model spaces, it is already challenging to
compute quadrupole moments in $p$-shell nuclei that are converged with
respect to the size of the HO model space, see
Refs.~\cite{cockrell2012,forssen2013,maris2013} for examples. This
motivates us to study the IR convergence for bound-state expectation
values of the quadrupole moment and for $E2$ transition matrix
elements between bound states.

This paper is organized as follows. We derive an extrapolation formula
for the deuteron's quadrupole moment and study our result for a toy
model and a realistic nucleon-nucleon interaction. We then generalize
the extrapolation formula to the general case where the bound-state
wave functions mixes partial waves with orbital angular momenta $l$
and $l+2$, respectively, or where the bound-state has a finite
$l>0$. Finally, we also derive an IR extrapolation formula for $E2$
transition matrix elements between bound states. We conclude with a
summary.


\section{Derivation}
\label{sec:derivation}
\subsection{Deuteron}

The deuteron is a spin-1 state 
\begin{equation}
\label{state}
\ket{\Psi} = \ket{\Psi_0} + \eta\ket{\Psi_2} ~,
\end{equation}
superposed of an $S$-state $\Psi_0$ and a $D$-state $\Psi_2$. The
$d$-state amplitude is denoted by $\eta$.  Without loss of generality
we focus on the state with maximum $J_z=1$ spin projection.  The wave
function for a state with orbital angular momentum $l$ is
\begin{equation}
\label{wf}
\Psi_l(r,\theta,\phi) = \mathcal{R}_l(r)\sum_{m, m_s} C_{l, m; 1, m_s}^{1, 1}
Y_{l,m}(\theta, \phi)\chi_{s, m_s} ~. 
\end{equation}
Here $\mathcal{R}_l(r)$ denotes the solution to the radial
Schr\"odinger equation. The orbital angular momentum, represented by
the spherical harmonics $Y_{lm}(\theta, \phi)$, and spin, represented
by the spinor $\chi_{s, m_s}$, are coupled to a total angular momentum
$J = 1$ by means of the Clebsch-Gordan coefficient $C_{l, m; 1,
  m_s}^{1, 1}$~\cite{Varshalovich:1988ye}.

For the computation of the IR correction of the quadrupole moment we
follow closely the corresponding derivation made in
Ref.~\cite{furnstahl2014} for the radius squared.  In a finite
oscillator basis with IR length scale $L$, the expectation value of
the quadrupole moment~(\ref{quad}) will differ from the infinite-space
result, and
\begin{equation}
Q_L = Q_\infty + \Delta Q_L ~.
\end{equation}
Here
\begin{equation}
\Delta Q_L = \frac{\braket{\Psi_L | \hat{Q} | \Psi_L}}{\braket{\Psi_L | \Psi_L}} - \frac{\braket{\Psi_\infty | \hat{Q} | \Psi_\infty}}{\braket{\Psi_\infty | \Psi_\infty}}~,
\label{deltaQ}
\end{equation}
defines the expressions for $Q_L$ and $Q_\infty$ such that any
$L$-independent terms will cancel in Eq.~\eqref{deltaQ}. The wave
functions $\Psi_L$ and $\Psi_\infty$ are the deuteron wave functions
in the finite and infinite oscillator spaces, respectively.

Using Eqs.~\eqref{state} and \eqref{wf}, four terms enter the
expectation value in the first term of Eq.~\eqref{deltaQ}, and

\begin{equation}
\begin{split}
\braket{\Psi_L | \hat{Q} | \Psi_L} = e\sqrt{\frac{\pi}{5}}
\int\limits_0^L \int\limits_0^\pi \int\limits_0^{2\pi} {\rm d}r r^2
{\rm d}\theta \sin\theta {\rm d}\phi \left[\mathcal{R}_{L,0}
  Y_{00}^*\chi_1^{\dagger} + \eta \mathcal{R}_{L,2}
\sum_{m, m_s} C_{2, m; 1, m_s}^{1, 1}Y_{2m}^*\chi_{m_s}^\dagger\right] \\
\times r^2 Y_{20} \left[\mathcal{R}_{L,0} Y_{00}\chi_1 + 
\eta \mathcal{R}_{L,2}\sum_{m, m_s} C_{2, m; 1, m_s}^{1, 1}Y_{2m}\chi_{m_s}\right]~.
\end{split}
\label{full_exp_value}
\end{equation}
The expectation value in the second term on the right-hand side of
Eq.~\eqref{deltaQ} is found by replacing $L$ by $\infty$. The $S$-$S$
term is zero, so we have only to consider the remaining $S$-$D$ mixing
terms and the $D$-$D$ term. Our interest is in the $L$-dependence of the
quadrupole moment which is contained entirely in the radial
integrations carried out in the first term on the right-hand side
of Eq.~\eqref{deltaQ}
\begin{equation}
\int\limits_0^L {\rm d}r r^4 \mathcal{R}_{L,0}(r)\mathcal{R}_{L,2}(r)~,
\end{equation}
and
\begin{equation}
\int\limits_0^L {\rm d}r r^4 \mathcal{R}_{L,2}(r)\mathcal{R}_{L,2}(r)~.
\end{equation}

We assume that the nuclear potential vanishes beyond $r=R$ and 
split the radial integration into two parts. In general,
\begin{equation}
\int_0^L {\rm d}r r^4 \mathcal{R}_{L,l_1}(r)\mathcal{R}_{L,l_2}(r) 
= \int_0^R {\rm d}r r^4 \mathcal{R}_{L,l_1}(r)\mathcal{R}_{L,l_2}(r) 
+ \int_R^L {\rm d}r r^4 \mathcal{R}_{L,l_1}(r)\mathcal{R}_{L,l_2}(r)~.
\label{splitradialint}
\end{equation}

The interior region, between $0$ and $R$, depends primarily on the
details of the interaction.  Around $E_\infty$ one assumes that the
radial wave function $\mathcal{R}_{L,l}$ in $L$-space can be expanded
in terms of the radial wave function in infinite space
$\mathcal{R}_{\infty,l}$ and a correction term, e.g. by using the
linear energy method~\cite{Djajaputra:2000aa}.  The resulting
$L$-dependence from the integration over the interior region scales as
$\mathcal{O}(L^0)e^{-2kL}$ \cite{furnstahl2014} and therefore does not
contribute to the dominant correction terms [the polynomial in $kL$ at
  $\mathcal{O}(e^{-2kL})$]. We therefore concentrate on the second
region between $R$ and $L$, and consider the integrals
\begin{equation}
\int\limits_R^L {\rm d}r r^4 \mathcal{R}_{L,0}(r)\mathcal{R}_{L,2}(r)~,
\label{radial_cross_term}
\end{equation}
and
\begin{equation}
\int\limits_R^L {\rm d}r r^4 \mathcal{R}_{L,2}(r)\mathcal{R}_{L,2}(r)~,
\end{equation}
in the region free from the potential. Here, the radial wave functions are
\begin{eqnarray}
\mathcal{R}_{L,0}(r) &=& h_0(ik_L r) + C_0 h_0(-ik_L r) \nonumber\\
&=& -\frac{e^{-k_L r}}{k_L r} + C_0 \frac{e^{k_L r}}{k_L r}~, \label{R0}
\end{eqnarray}
with 
\begin{equation}
C_0 = -\frac{h_0(ik_L L)}{h_0(-ik_L L)} =e^{-2k_L L}~, 
\end{equation}
and
\begin{eqnarray}
\mathcal{R}_{L,2}(r)&=&  h_2(ik_L r) + C_2 h_2(-ik_L r) \nonumber\\
		&=&  \frac{e^{-k_L r}}{(k_L r)^3}\left[(k_L r)^2 +
                  3k_L r + 3\right] 
- C_2\frac{e^{k_L r}}{(k_L r)^3}\left[(k_L r)^2 - 3k_L r + 3\right]~, \label{R2}
\end{eqnarray}
with
\begin{equation}
C_2 = -\frac{h_2(ik_L L)}{h_2(-ik_L L)} = e^{-2k_L L}\frac{(k_L L)^2 + 3k_L L + 3}{(k_L L)^2 - 3k_L L + 3}~.
\end{equation} 
The coefficients $C_0$ and $C_2$ are chosen such that the wave
function vanishes at $r = L$.

Finally, we sum the $S$-$S$ and $S$-$D$ terms and expand in
powers of $e^{-2kL}$. We consider the leading order (LO) term,
$Q_\infty$, and the next-to-leading order (NLO) term which contains a
polynomial in $kL$ times $e^{-2kL}$. We restrict our analysis to the
highest powers of $kL$ and arrive at
\begin{equation}
\label{master}
Q_L = Q_\infty - a(k_\infty L)^3\left(1+{\frac{d}{k_\infty L}}\right) e^{-2k_\infty L}~,
\end{equation}
with corrections of order $\mathcal{O}(k_\infty L e^{-2k_\infty
  L})$. Here $Q_\infty$, $a$, $d$, and $k_\infty$ can be treated as
fit parameters. Note that to LO, $k_L \approx k_\infty$, where (in the
two-nucleon system)
\begin{equation}
k_L = k_\infty - \gamma_\infty^2 e^{-2k_\infty L} + \mathcal{O}(e^{-4k_\infty L})~,
\end{equation}
and $\gamma_\infty$ is the asymptotic normalization coefficient~\cite{furnstahl2014}. The
LO term is all we need to determine the polynomial at
$\mathcal{O}(e^{-2k_\infty L})$ for $Q_L$.

\subsection{Generalized angular momentum states}
	
We can apply this reasoning to a system with an arbitrary mixture of
$l$ states, i.e. 
\begin{equation}
\ket{\Psi} = \ket{\Psi_{l_1}} + \eta \ket{\Psi_{l_2}}~.
\end{equation}
For simplicity we limit ourselves to LO and consider only the asymptotic form 
\begin{equation}
\label{asymp}
h_l(\rho) \rightarrow \frac{i}{\rho}e^{-i(\rho - \frac{l\pi}{2})}~,
\end{equation}
of the spherical Hankel functions at large $\rho$. As before, if we
consider $\pm i\rho$ (where $\rho = kr$) solutions and enforce the
boundary condition at $r = L$, we have for the radial behavior
\begin{equation}
\mathcal{R}_{L, l}(k_L r) = -\frac{1}{k_L r}e^{i\pi l/2}\left(e^{-k_L r} - e^{-2k_L L}e^{k_L r}\right)~.
\end{equation}

Computing the quadrupole moment expectation value again gives four
terms.  But when we consider the radial integrations
\begin{equation}
\label{rl1}		
\int\limits_R^L {\rm d}r r^4 \mathcal{R}^*_{L, l_1}(r)\mathcal{R}_{L, l_1}(r)~,
\end{equation}
\begin{equation}
\label{rl1rl2}
\int\limits_R^L {\rm d}r r^4 \mathcal{R}^*_{L, l_1(l_2)}(r)\mathcal{R}_{L, l_2(l_1)}(r)~,
\end{equation}
\begin{equation}
\label{rl2}
\int\limits_R^L {\rm d}r r^4 \mathcal{R}^*_{L, l_2}(r)\mathcal{R}_{L, l_2}(r)~,
\end{equation}
the $l$ dependence is either cancelled [as in Eq.~\eqref{rl1} and
  Eq.~\eqref{rl2}] or attributed to a phase [as in
  Eq.~\eqref{rl1rl2}], and they sum to give similar results.
Limiting ourselves to LO, we find
\begin{equation}
Q_L = Q_\infty - a(k_\infty L)^3 e^{-2k_\infty L}~.
\label{master_simp}
\end{equation}
We see that the general case agrees in LO with the particular
case~(\ref{master}) for the deuteron. Furthermore,
Eq.~(\ref{master_simp}) also applies to quadrupole expectation values
of bound-states with finite orbital angular momentum $l>0$ but no
mixing of partial waves. This makes Eq.~\eqref{master_simp} the main
result of this Subsection.  Higher-order corrections depend on orbital
angular momenta involved in the particular case under consideration.

\subsection{Electric quadrupole transitions}

The quadrupole moment operator also describes electric quadrupole
($E2$) transitions. If we consider a simple model where the initial
state is a pure $D$-wave state and the final state is a pure $S$-wave
state, the amplitude for such a transition is
\begin{equation}
\mathcal{A} = \braket{\Psi_0 | \hat{Q} | \Psi_2}~.
\end{equation}
As before, computing such an amplitude in a truncated basis
effectively imposes a Dirichlet boundary condition on the wave
functions. Likewise, we can describe the amplitude in the truncated
basis ($\mathcal{A}_L$) as the amplitude in the infinite basis
($\mathcal{A}_\infty$) plus a correction term.
\begin{equation}
\mathcal{A}_L = \mathcal{A}_\infty + \Delta \mathcal{A}_L~,
\end{equation}
where 
\begin{eqnarray}
\mathcal{A}_L & \equiv & \braket{\Psi_{L, 0} | \hat{Q} | \Psi_{L, 2}}~, \\
\mathcal{A}_\infty & \equiv & \braket{\Psi_{\infty, 0} | \hat{Q} | \Psi_{\infty, 2}}~,
\end{eqnarray}
and we seek to compute $\Delta \mathcal{A}_L$. We note that the
    bound-state momentum $k_l$ depends on the state $\Psi_l$. With
$\Psi_l$ from Eq. \eqref{wf}, and the radial wave functions from
Eqs. \eqref{R0} and \eqref{R2} for $\Psi_{L, 0}$ and $\Psi_{L, 2}$,
respectively, we can easily derive an expression for
$\mathcal{A}_L$. Essentially, we need to evaluate
Eq. \eqref{radial_cross_term} for states with different angular
momenta (or different $k_l$ values). While the procedure is similar
to the calculation of quadrupole moments, the result is somewhat
more complex. We obtain (written explicitly as a function of $L$)
\begin{equation}
\mathcal{A}_L = \mathcal{A}_{\infty} + a_0 \left[1 + 
\frac{a_1}{k_2 L} + \mathcal{O}\left(\frac{1}{(k_2 L)^2}\right)\right]e^{-2 k_2 L}~,
\label{E2convergence}
\end{equation}
where terms of $\mathcal{O}[(k_0 + k_2)L e^{-(k_0 + k_2)L}]$ and higher have been
dropped. Here, $k_0$ and $k_2$ represent the $S$- and $D$-wave binding
momenta, respectively (as $k_\infty$ previously represented the
separation energy in the case of the deuteron), and the constants
$a_0$ and $a_1$ are fit parameters.

In general, $E2$ transitions might occur between any states of
identical parities whose angular momenta differ by at most two units.
Employing the asymptotic form~(\ref{asymp}), we find that the
transition between bound states with angular momenta $l_1$ and $l_2$
and bound-state momenta $k_1$ and $k_2$, respectively extrapolates as
\begin{equation}
\label{trans}
\mathcal{A}_L = \mathcal{A}_{\infty} + a_0 e^{-2 k_< L}~. 
\end{equation}
Here, $k_< \equiv \min{(k_1,k_2)}$. The general LO
formula~(\ref{trans}) is the main result of this Subsection. One might
be surprised that the LO formula~(\ref{trans}) for $E2$ transitions
differs from the LO formula~(\ref{master_simp}) for expectation values
by the absence of the factor proportional to $L^3$. Inspection shows
that the limit $k_1\to k_2$ is interesting because terms with
prefactors involving $(k_1-k_2)^{-3}$ become proportional to $L^3$ in
this limit.


\section{Results}
\label{sec:results}
\subsection{Quadrupole moment in a square-well model}
As a first test of our analysis, we use a toy model for the deuteron consisting of a
square-well potential for the central and a tensor interaction, $V =
V_{\rm sq} + V_{\rm T}$, with
\begin{equation}
V_{\rm sq} = -V_0 \Theta(R - r)~,
\end{equation}
and
\begin{equation}
V_{\rm T} = \alpha V_{\rm sq} S_{12}~.		
\end{equation}
Here
\begin{eqnarray} 
S_{12} & = & \vec{Y_2}\cdot \vec{X_2} \nonumber \\
& = & \sum_\mu (-1)^\mu Y_{2,\mu} X_{2,-\mu}~,
\end{eqnarray}
consists of the rank-two tensor $\vec{Y}_2$ with components $Y_{2\mu}$, and
\begin{eqnarray} \vec{X_2}_\mu &\equiv&
\left(\vec{\sigma}\times\vec{\sigma}\right)^{(2)}_\mu \nonumber\\ 
&=&\sum_{m_s, m'_s} C^{2,\mu}_{1,m_s;1,m'_s}\sigma_{m_s}\sigma_{m'_s}~,
\end{eqnarray} 
is the rank-two spherical tensor obtained from coupling two spins.

We use units such that $\hbar=1$, $\mu=1$, and $R=1$. For the model
parameters we set $V_0 = 1.83$ (in units of $(\mu R^2)^{-1}$), and
$\alpha=0.5$.  This yields a $d$-state probability of about 4.1\%, a
ground-state energy of about $E\approx -0.41$ (in units of $(\mu
R^2)^{-1}$), and a squared radius of about $\langle \hat{r}^2\rangle\approx
0.36$ (in units of $R^2$).

\begin{center}
\begin{figure}
\includegraphics{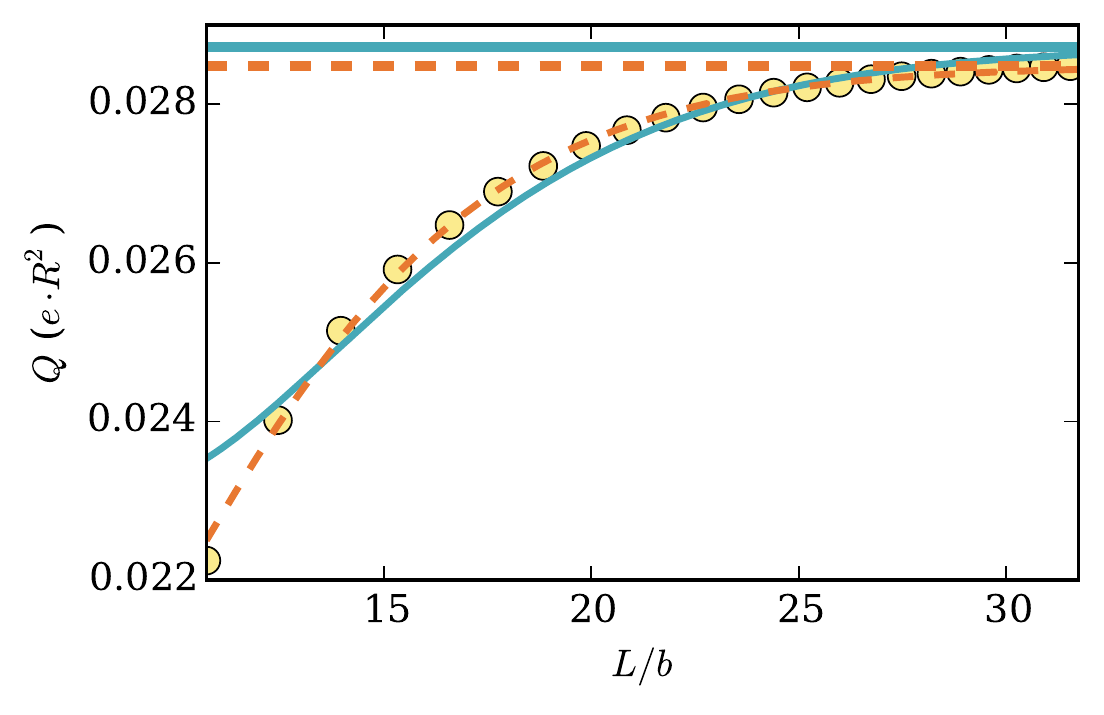}
\caption{(color online) Convergence of the quadrupole moment for the
  deuteron modeled by a simple square well potential as a function of
  the infrared length $L$ (given in units of the oscillator length
  $b$).  The yellow circles are the data from the computation. The
  thin, blue, solid line is the fit to Eq. \eqref{master_simp}. The
  thin, orange dashes are the fit to Eq. \eqref{master}. The thick,
  blue, solid line is $Q_\infty$ from the fit to
  Eq. \eqref{master_simp}. The thick, orange dashes are $Q_\infty$
  from the fit to Eq. \eqref{master}.}
\label{sqwell_qmo}
\end{figure}
\end{center}

We perform the diagonalization in the HO basis and compute the
quadrupole moment for an increasing number $N$ of oscillator
shells. We choose $\Omega = 28$ (in units of $(\mu R^2)^{-1}$)
yielding an oscillator length $b = 1 / \sqrt{28}$ (in units of $R$)
and $L = \sqrt{2(N + 3/2 + 2)}b$.  The results are shown in
Fig.~\ref{sqwell_qmo}. The oscillator frequency is chosen such that
$b\ll R$; the smallest length scale from the basis scales as
$b/\sqrt{N}$ and is thus adequate for a numerical diagonalization.

For the quadrupole extrapolation, we first extrapolate the
energy~(\ref{extraE}) and obtain the bound-state momentum $k_\infty$
from $E_\infty\equiv -(\hbar k_\infty)^2/(2\mu)$.  The
expression in Eq.~(\ref{master}) for the quadrupole extrapolation can be fit
in several ways.  We will first consider only the dominant term,
$(k_\infty L)^3$, and fit to Eq.~\eqref{master_simp}, treating $a$ and
$Q_\infty$ as fit parameters. The extrapolated $Q_\infty$ value is
within 1\% of the maximum $Q_\infty$ value ($N_{\rm{max}}$ = 500) when
fitting to values of $L$ as low as $L = 2.0$. Even fitting to the data
within a short range ($3.0 \le L \le 3.5$) provides an asymptotic
value within a percent of the value calculated by fitting to the
largest $L$ value. We also fit to Eq.~\eqref{master}, as shown in
Fig.~\ref{sqwell_qmo}. Including the extra fit parameter, $d$,
does not impact the extrapolated value significantly, though it can
influence the ranges of data over which we can accurately fit.

\subsection{Realistic deuteron quadrupole moment}

\begin{center}
\begin{figure}
\includegraphics{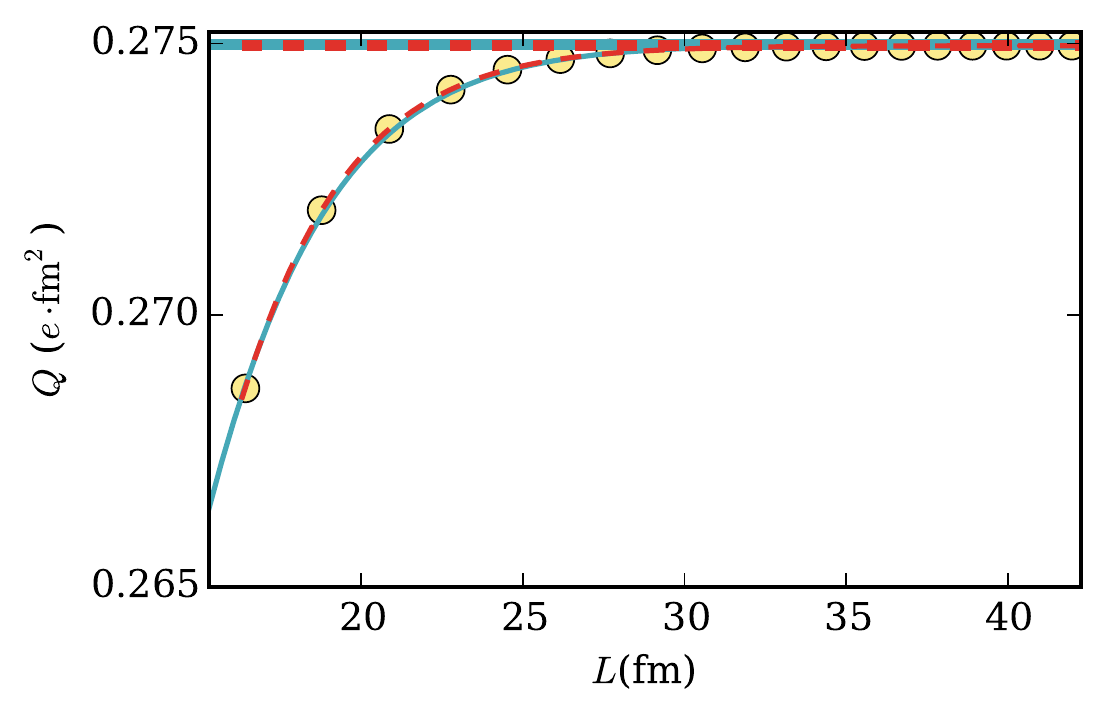}
\caption{(color online) Extrapolation of the deuteron quadrupole
  moment computed from a chiral potential. The yellow circles are the data
  from the calculation. The thin, blue, solid line is the fit to
  Eq. \eqref{master_simp}. The thick, blue, solid line is $Q_\infty$ from the fit
  to Eq. \eqref{master_simp}. The thin, red dashes are the fit to
  Eq. \eqref{master}, and the thick, red dashes are $Q_\infty$ from the fit to
  Eq. \eqref{master}.}
\label{dqmo}
\end{figure}
\end{center}

For an accurate model of the deuteron, we used the interaction from
chiral effective field theory (EFT) as described in
Ref.~\cite{machleidt2011}. For the quadrupole fits, we take the
bound-state momentum $k_\infty$ from the known binding energy of the
deuteron for this interaction.  Fits of the quadrupole moment to
Eqs.~\eqref{master} and \eqref{master_simp} yield virtually identical
results and are shown in Fig.~\ref{dqmo} and Table~\ref{tab1}.

\begin{center}
\begin{table}
\begin{tabular*}{0.4\textwidth}{@{\extracolsep{\fill}} c c c c}
\hline
$L_{\rm{min}}$ & $L_{\rm{max}}$ & $Q_\infty^{\eqref{master_simp}}$ & $Q_\infty^{\eqref{master}}$\\ 
\hline
20	&	42.40	&	0.2750	&	0.2750\\ 
\hline
15	&	42.40	&	0.2750	&	0.2750\\ 
\hline
10	&	42.40	&	0.2750	&	0.2750\\ 
\hline
10	&	20		&	0.2768	&	0.2740\\
\hline
20	&	30		&	0.2750	& 	0.2750\\
\hline
\end{tabular*}
\caption{$L_{\rm{min}}$ and $L_{\rm{max}}$ (in fm) are the range over
  which the data is fit. $Q_\infty^{\eqref{master_simp}}$ and
  $Q_\infty^{\eqref{master}}$ are the extrapolated quadrupole moments
  (in $e\cdot$~fm$^2$) when fitting the Eq. \eqref{master_simp} and
  Eq. \eqref{master} respectively.}
\label{tab1}
\end{table}
\end{center}

To illustrate the robustness of the extrapolations, we employ
different bound-state momenta, namely $k_\infty$ from the separation
energy, $k_E$ from a fit of the extrapolation~(\ref{extraE}) to the
ground-state energy, and $k_Q$ from a fit of the quadrupole
extrapolation formulas~(\ref{master}) and (\ref{master_simp}),
respectively, in the extrapolation formulas.  
While the values for the bound-state
momenta can differ by as much as 20\%, the extrapolated quadrupole
moments $Q_\infty$ differ by only about 1\%.

\begin{center}
\begin{figure}
\includegraphics{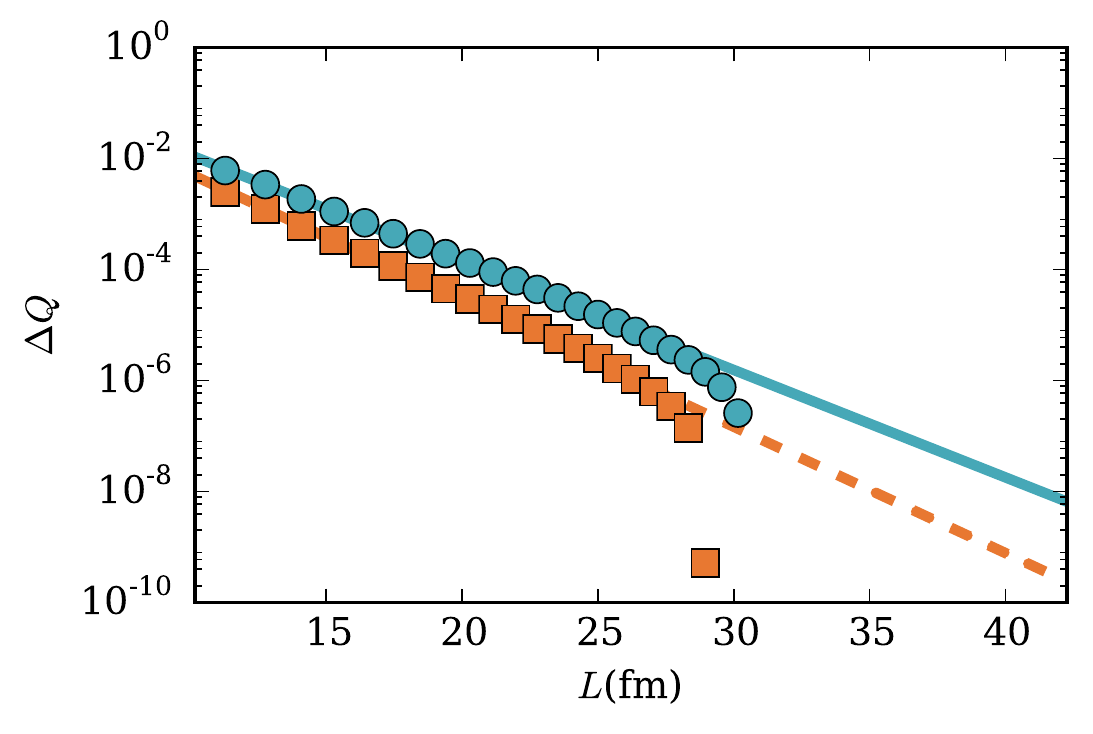}
\caption{Difference between the $L$-dependent quadrupole moment and
  its asymptotic value from an extrapolation based on
  Eq.~\eqref{master_simp} (blue circles are data and the solid blue
  line is the fit) and from an extrapolation based on the more precise
  Eq.~\eqref{master} (orange squares are data and the dashed orange
  line is the fit).}
\label{logdelta}
\end{figure}
\end{center}

Let us compare the differences between fitting to Eq.~\eqref{master}
and Eq.~\eqref{master_simp}. Figure~\ref{logdelta} shows that fitting
to the more precise Eq.~\eqref{master} improves the convergence
consistently. We note, however, that the difference in the extrapolated
values is almost negligible.

\begin{center}
\begin{figure}
\includegraphics{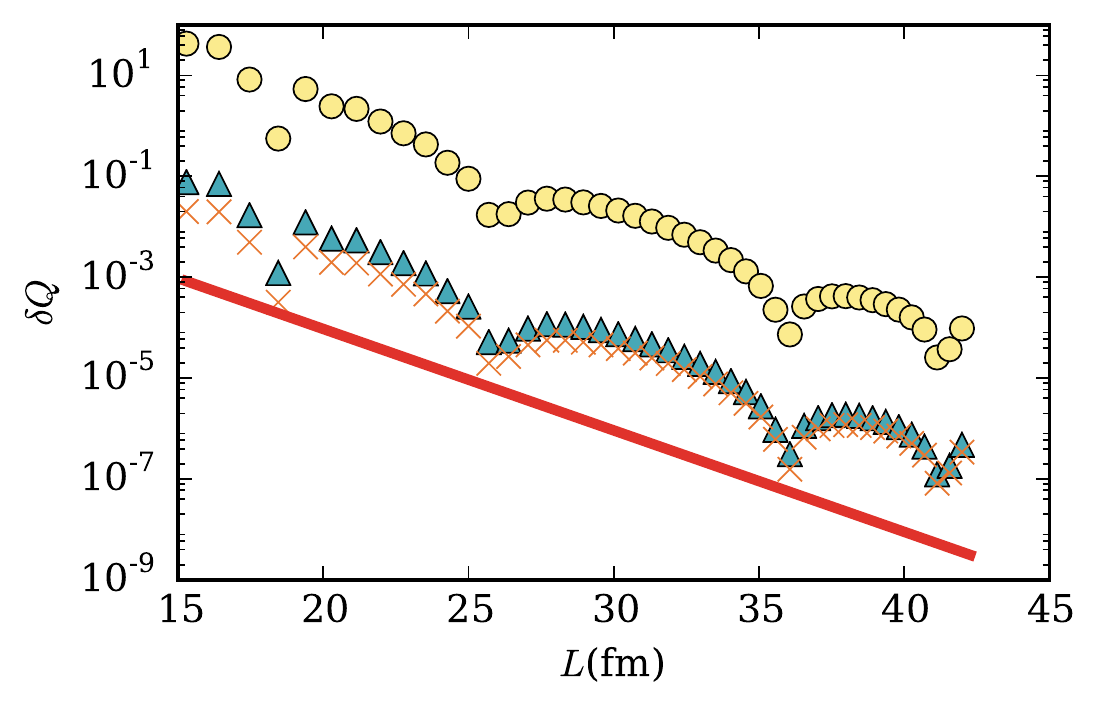}
\caption{(color online) The yellow circles represent $\delta Q_1$ as
  defined by Eq. \eqref{residual}.  The blue triangles represent
  $\delta Q_0$. The orange crosses represent $\delta Q_{-1}$. The red
  line is proportional to $e^{-2k_\infty L}$.}
\label{residual_plot}
\end{figure}
\end{center}

One might also try to extract higher-order corrections to the
quadrupole moment that are smaller than $\mathcal{O}((k_\infty L)^2
e^{-2k_\infty L})$. These are terms proportional to $(k_\infty L)^m
e^{-2k_\infty L}$ with $m\le 1$, and terms proportional to
$e^{-4k_\infty L}$. We define (with $Q(L)\equiv Q_L$)
\begin{equation}
\delta Q_m = \frac{Q_{\rm{calculated}} -
Q_{m+1}(L)}{c_m (k_\infty L)^m}~,
\label{residual}
\end{equation}
where 
\begin{equation}
Q_m(L) = \sum_{n = m}^3 c_n (k_\infty L)^n e^{-2 k_\infty L}~,
\end{equation}
is the data reproduced with the fit parameters (represented by $c_n$
and $c_m$). If we plot $\delta Q$ alongside what we expect
analytically, we ought to be able to establish trends for the higher
order corrections. The results are shown in
Fig.~$\ref{residual_plot}$. The overall slopes of the corrections
match well with the predicted slope, and as each lower order of
$(k_\infty L)$ is included, the data approaches the $e^{-2k_\infty L}$
line as expected, supporting the validity of our analysis.
		
\subsection{Electric quadrupole transitions}

\begin{center}
\begin{figure}
\includegraphics{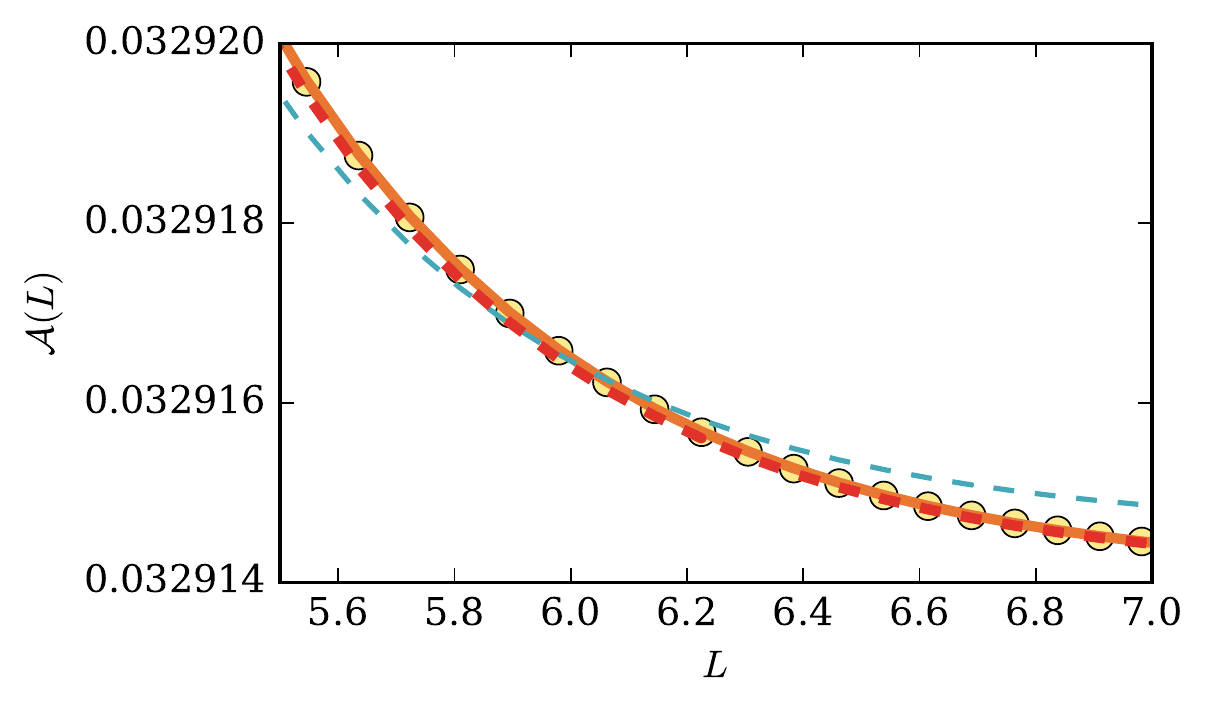}
\caption{(color online) Convergence of the $E2$ transition amplitude
  in $e \cdot fm^2$ as a function of $L$ in units of the $R$ (the
  range of the interaction) for a simple Gaussian well model. The
  yellow dots are the results of the numerical calculation. The
  thin, light blue dashes are the fit to the leading order result
  in Eq.~\eqref{E2convergence}. The thick, red dashes correspond to a fit
  including the $(1 / k_2 L)^2$ term in Eq.~\eqref{E2convergence}. 
  And the solid orange line includes the $(1 / k_2 L)^3$ term in
  the fit.}
\label{transition}
\end{figure}
\end{center}

To test our result~(\ref{E2convergence}) for $E2$ transitions, we
employ a Hamiltonian with a Gaussian well potential 
\begin{equation}
V(r) = -V_0 e^{-\left({r\over R}\right)^2}~,
\end{equation}
that is deep enough to contain a bound $D$-wave state as well as the
ground $S$-wave state. As parameters we choose $R=1$ and $V_0=15$ (in
units of $(\mu R^2)^{-1}$). Recall that the bound-state momenta of the
$S$- and $D$-states are $k_0$ and $k_2$, respectively. Because $k_2 <
k_0$, the dominant correction contains the exponential $e^{-2k_2 L}$.
Below, we consider three different orders of the polynomial in $k_2 L$
preceding the exponential $e^{-2k_2 L}$ that governs the correction
term. From Fig.~\ref{transition} we can see that the data is more
accurately described as increasing powers of $1 /(k_2 L)$ are
considered. However, over large ranges of $k_2 L$, which we are able
to take advantage of in the simple model presented here, the leading
order result can be sufficient to obtain accurate asymptotic
values. Figure~\ref{transition} highlights a small region where the
differences in the fitting can be seen. Lower and higher values of $L$
do not show significant differences.

\begin{center}
\begin{figure}
\includegraphics{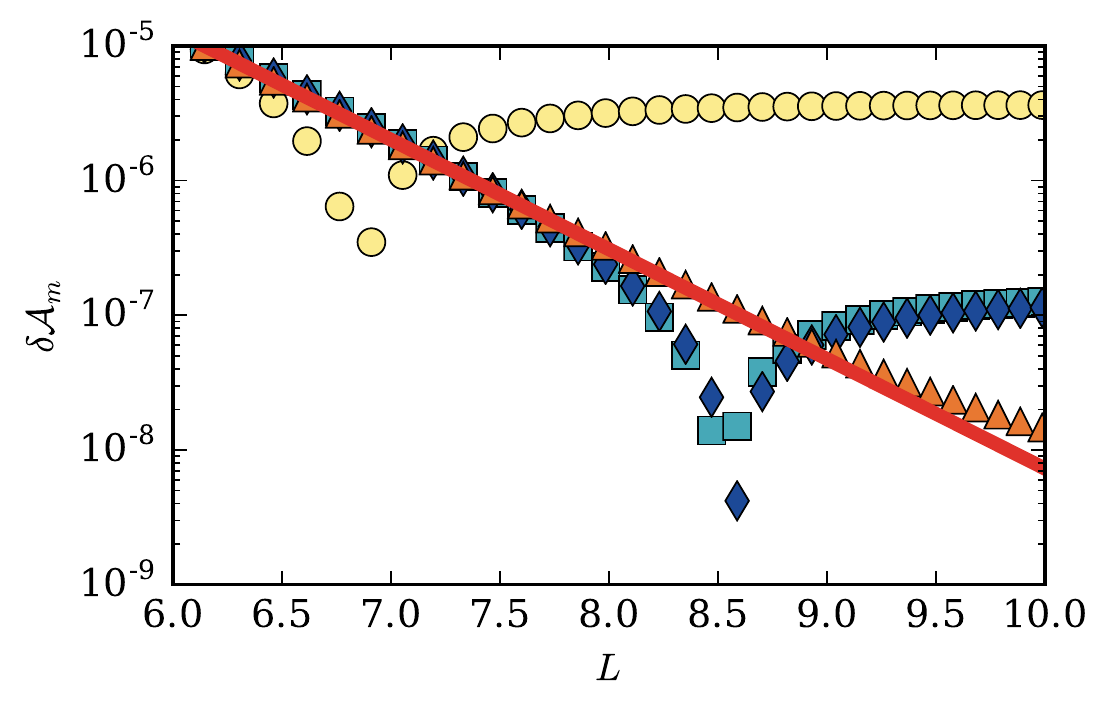}
\caption{(color online) The residual transition amplitude from fitting
  to increasing powers of $(1 / k_2 L)$ in Eq.~\eqref{E2convergence}. The
  yellow dots represent a leading order fit ($m = 0$), the light blue
  squares include also the $(1 / k_2 L)$ term ($m = -1$), the dark
  blue diamonds include also the $(1 / k_2 L)^2$ term ($m = -2$), and
  the orange triangles include also the $(1 / k_2 L)^3$ term
  ($m = -3$). The red, solid line represents $e^{-2 k_2 L}$ where
  $k_2$ is determined from a $d$-wave energy fit.}
\label{transitionlog}
\end{figure}
\end{center}

The log-scale plot, shown in Fig.~\ref{transitionlog}, reveals the
differences between the fits and, more importantly, the improvement as
higher orders of $(1 / k_2 L)$ are included. Here, we plot the
residual transition amplitude, {\it i.e.} the difference between the values
calculated in the truncated basis ($\mathcal{A}_{\rm calculated}$) and the
values reproduced by the fit parameters ($A_\infty$ and $c_n$).
We define 
\begin{equation}
 \delta \mathcal{A}_m =
\frac{\mathcal{A}_{\rm calculated} - \mathcal{A}_\infty}{\sum_{n = m}^0 c_n
  (k_2 L)^n}~,
\end{equation}
where $m < 0$, and we plot the result in Fig.~\ref{transitionlog} as a
measure of how well the fit describes the data for different $m$
values. Little improvement comes from the $m=-2$ term due to its small
coefficient. Most importantly, we can see that the deviation of the
data from the expected behavior happens at larger and larger $L$
values as more terms in the polynomial factor are included.

\section{Summary}
\label{sec:summary}	
We derived IR extrapolation formulas for bound-state quadrupole
moments and for quadrupole transitions between bound states in finite
oscillator spaces. For two-body systems, the extrapolations are of the
form $(k L)^n e^{-2 k L}$ with $k$ denoting a bound-state momentum,
$L$ the IR length of the oscillator basis, and an integer $n$. We
successfully tested the extrapolation formulas (and higher-order
corrections) in simple potential models and for a realistic deuteron
computed with interactions from chiral EFT. It would be interesting to
probe and use these formulas in {\it ab initio} computations of finite
nuclei.  Our results for quadrupole transitions between bound states
should also hold for transitions from bound states into narrow states
close to the threshold. It would also be interesting to work out
extrapolation formulas for transitions into arbitrary continuum states
in the future.
	
\begin{acknowledgments}
  We thank A. Ekstr\"om for providing us with matrix elements. This
  work was supported in parts by the U.S. Department of Energy, Office
  of Science, Office of Nuclear Physics under Grant
  No. DEFG02-96ER40963 (University of Tennessee) and Contract
  No. DE-AC05-00OR22725 (Oak Ridge National Laboratory), the National
  Science Foundation under Grant No. PHY-1516077, and by the US-Israel
  Binational Science Foundation under Grant No. 2012212.
\end{acknowledgments}

%

\end{document}